# One Signal-Noise Separation based Wiener Filter for Magnetogastrogram

Hua Li

*Abstract*—Magnetogastrogram (MGG) signal frequency is about 0.05 Hz, the low-frequency environmental noise interference is serious and can be several times stronger in magnitude than the signals of interest and may severely impede the extraction of relevant information. Wiener filter is one classic denoising solution for biomagnetic applications. Since the reference channels are usually placed not far enough from the biomagnetic sources under test, they will inevitably detect the signals and the Wiener filters may produce ill-conditioned solutions. Considering the solutions to improve the signal-to-noise ratio (SNR) of Wiener filter output, there are few methods to separate the signals from the noises of the reference signal at the filter input. In this paper, a new signal processing framework called signal-noise separation based Wiener filter (SNSWF) is proposed that it separates the main noise as the input signal of the filter to improve the output SNR of Wiener filter. The filter was successfully applied to the noise suppression for MGG signal detection. Using the SNSWF, the filter SNR is 16.7 dB better than the classic Wiener filter.

## I. INTRODUCTION

MAGNETOGASTROGRAM is a noninvasive technique to provide magnetic information about the gastric electrophysiological activities with high resolution SQUID (Superconducting QUantum Interference Device) sensors [1], [2]. It records the magnetic field generated by the gastric basic electrical rhythm (BER) and is an effective technique for detecting gastric function. The MGG signal magnetic strength is on the pT scale and the slow-wave frequency (SWF) is about 3 ± 0.5 cpm (cycles per minute) [3], [4]. Since the signal frequency is about 0.05 Hz, the low-frequency environmental noise interference is serious and can be several times stronger in magnitude than the signals of interest and may severely impede the extraction of relevant information.

In order to acquire the interest biomagnetic sources from the environmental noise sources and the artifactual sources, linear methods and nonlinear methods are involved [5-9]. Besides the wiener filter, spatial filtering such as signal space projection (SSP), principal component analysis (PCA), common spatial subspace decomposition (CSSD), and independent component analysis (ICA) are widely used as linear methods. The nonlinear methods rely on nonlinear kernel-based algorithms such as reproducing kernel Hilbert spaces (RKHSs), kernel blind source separation (KBSS).

Wiener filter is one widely used denoising solution [10], especially for biomagnetic applications [11], [12]. It often transforms a linear combination of instantaneous values along with its delays as an input signal to closely match a desired or target signal by a linear time-invariant (LTI) filter. The destination of a Wiener filter is to minimize the mean-square error between the estimated signal sequence and the desired signal sequence. In biomagnetic noise canceling problems, the input signals are from several reference channels which can detect the environmental noise. The output of the signal channels can be used as the desired signal. Since the reference channels are usually placed not far enough from the biomagnetic sources under test, they will inevitably detect the signals and the Wiener filters may produce ill-conditioned solutions. There are few methods to improve the signal-to-noise ratio (SNR) of Wiener filter output by separating the signals from the noises of the reference signal at the filter input. The major contributions of this work can be highlighted as follows:

- A new signal processing framework to improve the output signal-to-noise ratio of Wiener filter is proposed: minimize the source signal in the reference signal, and separate the main noise as the input signal of the filter to improve the output SNR of Wiener filter. The main operation process is as follows: First, before the reference signal enters the filter, a signal-to-noise separation step is carried out based on second order blind identification (SOBI) method. Second, the power spectrum analysis of the separated components is carried out which use the Autoregressive (AR) spectral analysis methods. Three, through the main power spectrum distribution range, the noise components are retained as the input signals to the greatest extent, rather than selecting the signal components into the filter.
- The SNSWF was successfully applied to the noise suppression for human MGG signal detection.

This paper is organized as follows. In the next section, we present problem and derive the research motivation. In Section III, some necessary prerequisites on the theory of signal-noise separation based Wiener Filter are given. Signal-noise separation based Wiener Filter with adjustable cutoff frequency are established. Experimental results in MGG

This work was supported in part by the National Key R&D Program of China: BTIT under Grant 2022YFF1202800)."

Hua Li is with the National Key Laboratory of Materials for Integrated Circuits, Shanghai Institute of Microsystem and Information Technology, Chinese Academy of Sciences, 865 Changning Road, Shanghai 200050 China (e-mail: leehua@ shu.edu.cn).

signals denoising are presented in Section IV. Some concluding remarks are given in Section V.

## II. PROBLEM FORMULATION AND MOTIVATION

### A. The time-domain expression of the Wiener filter

The canonical form of the Wiener filter is proposed in Fig. 1.

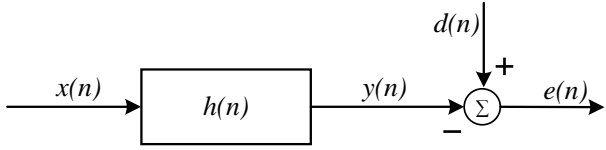

**Fig. 1.** The input-output relationship of the Wiener filter.

Where x(n) is the signal observed in reality, and after filtering by h(n), we get y(n).

$$y(n) = \sum_{m=0}^{+\infty} h(m)x(n-m). \quad (1)$$

The estimated y(n) signal from the block diagram of Figure 1 cannot be exactly the same as the desired useful signal d(n), where e(n) is used to represent the error between the true value and the estimated value.

$$e(n) = d(n) - y(n). \quad (2)$$

The error criterion of the Wiener filter is the minimum mean squared error.

$$E\left[e^2(n)\right] = E\left[(d(n) - y(n))^2\right]. \quad (3)$$

Let h(n) be physically achievable, i.e. the causal sequence:

$$h(n) = 0, \text{ when } n<0. \quad (4)$$

Therefore, it can be derived:

$$E\left[e^2(n)\right] = E\left[\left(d(n) - \sum_{m=0}^{+\infty} h(m)x(n-m)\right)^2\right]. \quad (5)$$

To minimize the mean squared error, take the deviation of the above equation for each h(m), m=0, 1, …, and equal to zero, to get:

$$2E\left[\left(d(n) - \sum_{m=0}^{+\infty} h(m)x(n-m)\right)x(n-k)\right] = 0$$

$$k=0,1,2\ldots \quad (6)$$

That is:

$$E[d(n)x(n-k)] = \sum_{m=0}^{+\infty} h(m) E[x(n-m)x(n-k)] \quad k \geq 0 \quad (7)$$

Expressing the above equation using the correlation function r gives the discrete form of the Wiener-Hoff equation:

$$r_{xd}(k) = \sum_{m=0}^{+\infty} h(m) r_{xx}(k-m) \quad k \geq 0 \quad (8)$$

The h(n) solved from the Wiener-Hoff equation is the best h(n), $h_{opt}(n)$ at the minimum mean squared error. Let h(n) be a causal sequence and can be approximated with a sequence of finite length (N point), then:

$$r_{xd}(k) = \sum_{m=0}^{N-1} h_{opt}(m) r_{xx}(k-m) \quad k=0,1,2, \ldots, N\text{-}1 \quad (9)$$

This gives N linear equations:

$$\begin{cases} k=0 & r_{xd}(0) = h(0)r_{xx}(0) + h(1)r_{xx}(1) + \cdots + h(N-1)r_{xx}(N-1) \\ k=1 & r_{xd}(1) = h(0)r_{xx}(1) + h(1)r_{xx}(1) + \cdots + h(N-1)r_{xx}(N-2) \\ \vdots & \vdots \\ k=N-1 & r_{xd}(N-1) = h(0)r_{xx}(N-1) + h(1)r_{xx}(N-2) + \cdots + h(N-1)r_{xx}(0) \end{cases} \quad (10)$$

Written in matrix form:

$$\begin{bmatrix} r_{xx}(0) & r_{xx}(1) & \cdots & r_{xx}(N-1) \\ r_{xx}(1) & r_{xx}(0) & \cdots & r_{xx}(N-2) \\ \vdots & \vdots & \cdots & \vdots \\ r_{xx}(N-1) & r_{xx}(N-2) & \cdots & r_{xx}(0) \end{bmatrix} \begin{bmatrix} h(0) \\ h(1) \\ \vdots \\ h(N-1) \end{bmatrix} = \begin{bmatrix} r_{xd}(0) \\ r_{xd}(1) \\ \vdots \\ r_{xd}(N-1) \end{bmatrix} \quad (11)$$

Simplified form:

$$R_{xx} H = R_{xd} \quad (12)$$

where H is the unit shock response to be solved; $R_{xd}$ are the cross-correlation sequences; $R_{xx}$ is the autocorrelation matrix. As long as $R_{xx}$ is non-singular, it can be solved to get H:

$$H = R_{xx}^{-1} R_{xd} \quad (13)$$

### B. Frequency-domain form of Wiener-Hoff equation

The Wiener – Hoff equation (8) can be written in convolutional form:

$$r_{xd}(k) = h(k) * r_{xx}(k) \quad (14)$$

Applying the Z-transformation to the above equation yields:

$$r_{xd}(z) = h(z) r_{xx}(z) \quad (15)$$

Where $r_{xx}(z)$ is the power-density spectrum of the input signal which is the Z transform of $r_{xx}(k)$. $r_{xd}(z)$ is the cross-power spectrum between the input signal and desired response which is the Z transform of $r_{xd}(k)$. h(z) is the transfer function of the Wiener filter:

$$h(z) = \frac{r_{xd}(z)}{r_{xx}(z)} \quad (16)$$

### C. SNR analysis of Wiener filters

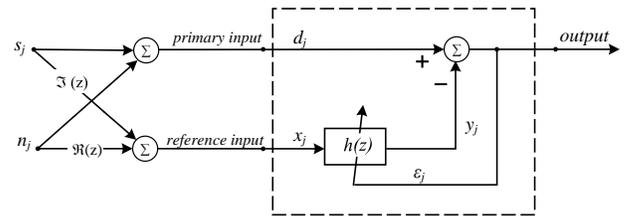

**Fig. 2.** Wiener filter with signal components in the reference input.

The spectrum of the noise n is $r_{nn}(z)$ and its transfer function is $\Re(z)$. The spectrum of the noise s is $r_{ss}(z)$ and its transfer function is $\Im(z)$, as shown in Fig. 2. The spectrum of the noise n arriving via $\Re(z)$ is $r_{nn}(z)|\Re(z)|^2$ and the spectrum of the signal s arriving via $\Im(z)$ is $r_{ss}(z)|\Im(z)|^2$. The spectrum of the reference input $x_j$ to the filter is thus

$$r_{xx}(z) = r_{ss}(z)|\Im(z)|^2 + r_{nn}(z)|\Re(z)|^2 \quad (17)$$

The cross-spectrum between the primary input $x_j$ and the reference input $d_j$ is similarly

$$r_{xd}(z) = r_{ss}(z)\Im(z^{-1}) + r_{nn}(z)\Re(z^{-1}) \quad (18)$$

The Wiener filter transfer function given by (16), is thus

$$h(z) = \frac{r_{ss}(z)\Im(z^{-1}) + r_{nn}(z)\Re(z^{-1})}{r_{ss}(z)|\Im(z)|^2 + r_{nn}(z)|\Re(z)|^2} \quad (19)$$

$\delta_{out}(z)$ is the signal-to-noise density ratio at the Wiener filter output. The transfer function of the propagation path from the signal input to the filter output is $1-\Im(z)h(z)$ and that of the path from the noise input to the filter output is $1-\Re(z)h(z)$. The spectrum of the signal component in the output is thus

$$r_{ss_{out}}(z) = r_{ss}(z)|1 - \Im(z)h(z)|^2$$
$$= r_{ss}(z)\left|\frac{[\Re(z) - \Im(z)]r_{nn}(z)\Re(z^{-1})}{r_{ss}(z)|\Im(z)|^2 + r_{nn}(z)|\Re(z)|^2}\right|^2 \quad (20)$$

and that of the noise component is

$$r_{nn_{out}}(z) = r_{nn}(z)|1 - \Re(z)h(z)|^2$$
$$= r_{nn}(z)\left|\frac{[\Im(z) - \Re(z)]r_{nn}(z)\Im(z^{-1})}{r_{ss}(z)|\Im(z)|^2 + r_{nn}(z)|\Re(z)|^2}\right|^2 \quad (21)$$

The $\delta_{out}(z)$ is thus

$$\delta_{out}(z) = \frac{r_{ss_{out}}(z)}{r_{nn_{out}}(z)} = \frac{r_{nn}(z)|\Re(z)|^2}{r_{ss}(z)|\Im(z)|^2} \quad (22)$$

The signal-to-noise density ratio of the reference input is $\delta_{ref}(z)$ and can be expressed as follows. The spectrum of the signal component in the reference input is thus

$$r_{ss_{ref}}(z) = r_{ss}(z)|\Im(z)|^2 \quad (23)$$

and that of the noise component is

$$r_{nn_{ref}}(z) = r_{nn}|\Re(z)|^2 \quad (24)$$

The signal-to-noise density ratio at reference input is

$$\delta_{ref}(z) = \frac{r_{ss}(z)|\Im(z)|^2}{r_{nn}(z)|\Re(z)|^2} \quad (25)$$

The signal-to-noise density ratio at filter output is

$$\delta_{out}(z) = \frac{1}{\delta_{ref}(z)} \quad (26)$$

It shows that the output $\delta_{out}(z)$ is the reciprocal of the reference input $\delta_{ref}(z)$. Therefore, if people want to improve the output SNR of the Wiener filter, the best way is to minimize the reference input SNR, which is also the motivation of our research work.

## III. SIGNAL-NOISE SEPARATION BASED WIENER FILTER

### A. Basic principle of signal-noise separation based wiener filter

In order to minimize the SNR of the reference input, we propose a new solution as shown in Fig. 3, which can be called signal-noise separation based Wiener filter (SNSWF).

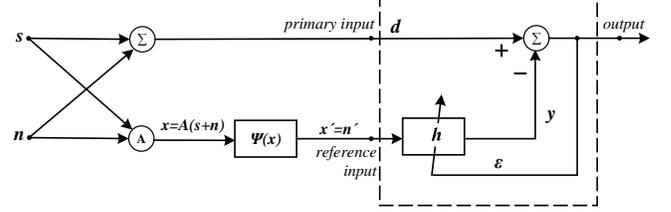

**Fig. 3.** The structure of the signal-noise separation based Wiener filter.

The reference channels detect the noise sources at the same time that the signal sources are also detected. Let us consider the reference signals **x** are instantaneous linear mixtures of the unknown signal sources **s** and the noise sources **n**. The mixing model is thus

$$\mathbf{x} = \mathbf{A}(\mathbf{s} + \mathbf{n}) \quad (27)$$

Where **A** is the mixing matrix. In order to separate the noise **n** and reduce the SNR of the reference signals **x**, it is necessary to find a suitable operating function $\psi$ to calculate the demixing matrix **W**.

$$\psi(\mathbf{x}) = \psi[\mathbf{A}(\mathbf{s}+\mathbf{n})] \quad (28)$$

$$\mathbf{W} = \mathbf{A}^{-1} \quad (29)$$

After demixing, the main noise component **n′** is selected as the filter reference input **x′**.

$$\mathbf{x'} = \mathbf{n} \quad (30)$$

For solving this problem, two assumptions are made: the mixing matrix A is nonsingular; the sources are spatially uncorrelated.

Several blind source separation algorithms such as PCA, ICA can be used as the operating function. In our work, we choose the second-order blind identification (SOBI) algorithm [13]. The implementation of the SOBI algorithm is thus

1) Estimate the sample covariance **C**(0) from data samples. Denote by $\lambda_1, \ldots, \lambda_n$, the n largest eigenvalues and $\mathbf{p}_1, \ldots, \mathbf{p}_n$, the corresponding eigenvectors of **C**(0).
2) An estimate $\sigma^2$ of the noise variance is the average of the smallest eigenvalues of **C**(0). The whitened signals are $\mathbf{z}(t)=[z_1(t), \ldots, z_n(t)]^T$, which are computed by $z_i(t)=(\lambda_1-\sigma^2)^{-(1/2)}\mathbf{p}_i^*\mathbf{x}(t)$ for $1 \leq i \leq n$. This is equivalent to forming a whitening matrix by

$$\Lambda = \left[(\lambda_1-\sigma^2)^{-1/2} p_1, \cdots, (\lambda_n-\sigma^2)^{-1/2} p_n\right]^H.$$

   Superscript * denotes the conjugate transpose of a vector and superscript H denotes the complex conjugate transpose of a matrix.
3) Form sample estimates **C**($\tau$) by computing the sample covariance matrices of **z**(t) for a fixed set of time lags $\tau \in \{\tau_j \mid j=1, \ldots, K\}$.
4) A unitary matrix **U** is then obtained as joint diagonalizer of the set $\{\mathbf{C}(\tau_j) \mid j=1, \ldots, K\}$.

5) The source signals are estimated as $\hat{\mathbf{x}}(t) = \mathbf{U}^H \mathbf{\Lambda} \mathbf{x}(t)$, and the mixing matrix $\mathbf{A}$ is estimated as $\mathbf{A} = \mathbf{\Lambda}^{\#} \mathbf{U}$. Superscript # denotes the Moore–Penrose pseudoinverse.

After the SOBI operation, the Autoregressive (AR) spectral analysis of the separated components is carried out. Through the main power spectrum distribution range, the noise components are retained as the input signals in order to minimize the SNR of the reference input.

## IV. THE SNSWF SIMULATION RESULTS

### A. The simulation design and signal-noise separation results

In MGG detections, we use the second-order axial gradiometer as the signal channel ($s_g$) with baseline of 5 cm and diameter of 18 mm. Three magnetometers and five one-order tensor gradiometers are used as reference channels (R1, …, R8) which can detect the environmental noise [14]. Each signal and reference sensor are coupled to one SQUID. SQUIDs are immersed in liquid helium at the temperature of 269 C° that convert the magnetic flux to voltage. In order to ensure the effectiveness of the simulation, the outputs of the reference channels and the signal channel in the real environment without the measured person were collected as the background noises of the simulation. Based on these background noises, single-frequency noise and signal are added to verify SNSWF performance.

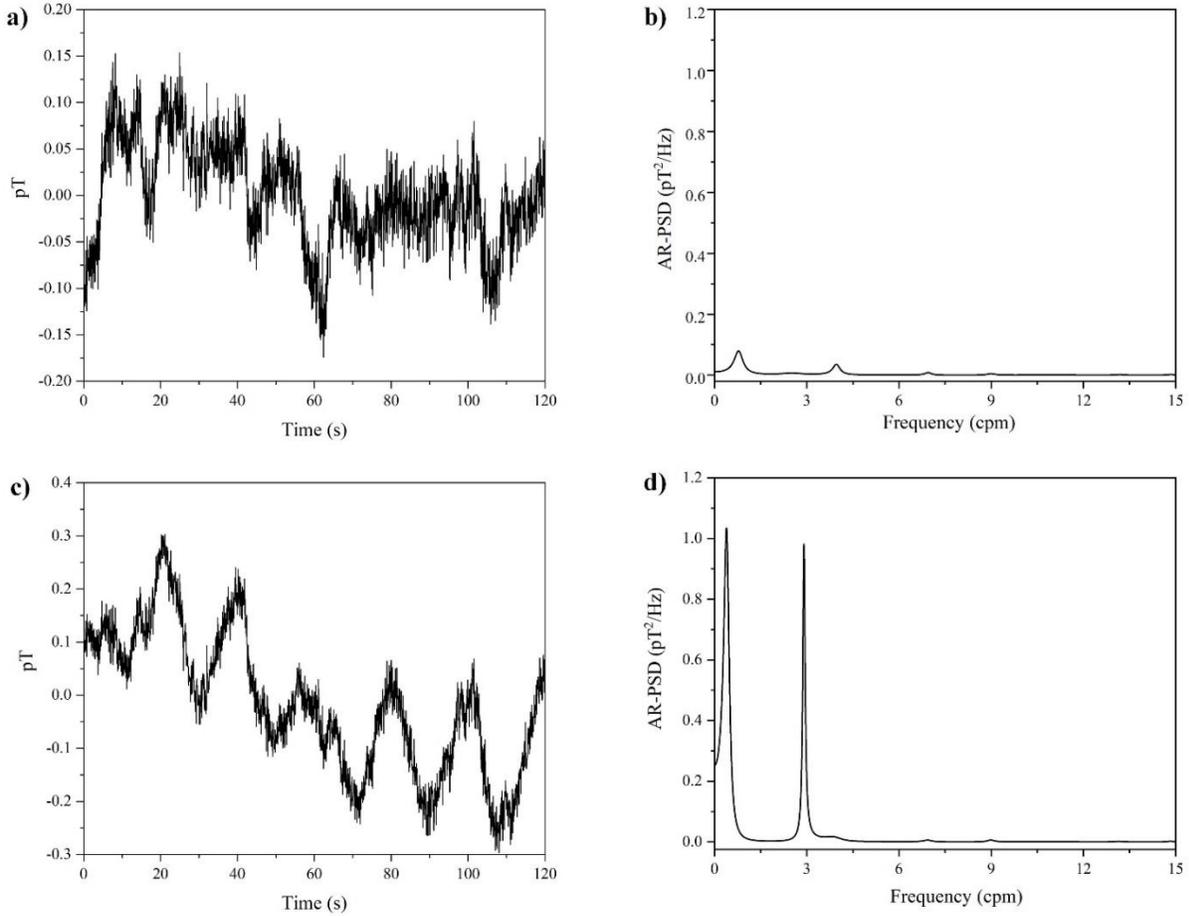

**Fig. 4.** (a) The output of second-order axial gradiometer without tested person. (b) The AR-PSD for the signal channel output. (c) The simulated signal gradiometer ($s_{sg}$) output. (d) The AR-PSD of $s_{sg}$.

Therefore, single-frequency signal $s_s(t)$ and noise $n_s(t)$ are used as the simulation signal and simulation noise $x_s(t)$. Considering about the MGG slow-wave frequency is about 3 cpm, we set the signal frequency $f_s$ is 3 cpm and one noise frequency $f_n$ is 0.3 cpm. The simulated signal gradiometer ($s_{sg}$) outputs are given by the following expression:

$$x_s(t) = s_s(t) + n_s(t) = \cos(2\pi f_s t) + \cos(2\pi f_n t) \quad (31)$$

$$s_{sg} = s_g + 0.1 x_s(t) \quad (32)$$

Fig. 4(a) is the output of the signal channel with amplitude of ± 0.2 pT. The AR-PSD of this gradiometer output is shown in Fig. 4(b) which has two stronger peaks at 0.8 cpm and 3.9 cpm.

Fig. 4(c) is simulated output of the signal channel. The 3 cpm signal and 0.3 cpm noise can be seen from the Fig. 4(c).

Fig. 4(d) is corresponding AR-PSD of the signal channel and the two peaks frequency are 0.39 cpm (1.03 pT$^2$/Hz) and 2.9 cpm (0.98 pT$^2$/Hz). Due to the influence of the added ambient noise, the frequency of the simulated signal and noise have an error about 0.1 cpm from the set frequency value. The SNR can be defined as follow:

$$SNR = 20\log(\frac{signal_{AR-PSD}}{noise_{AR-PSD}}) \quad (33)$$

The SNR of $s_{sg}$ is about -0.43 dB.

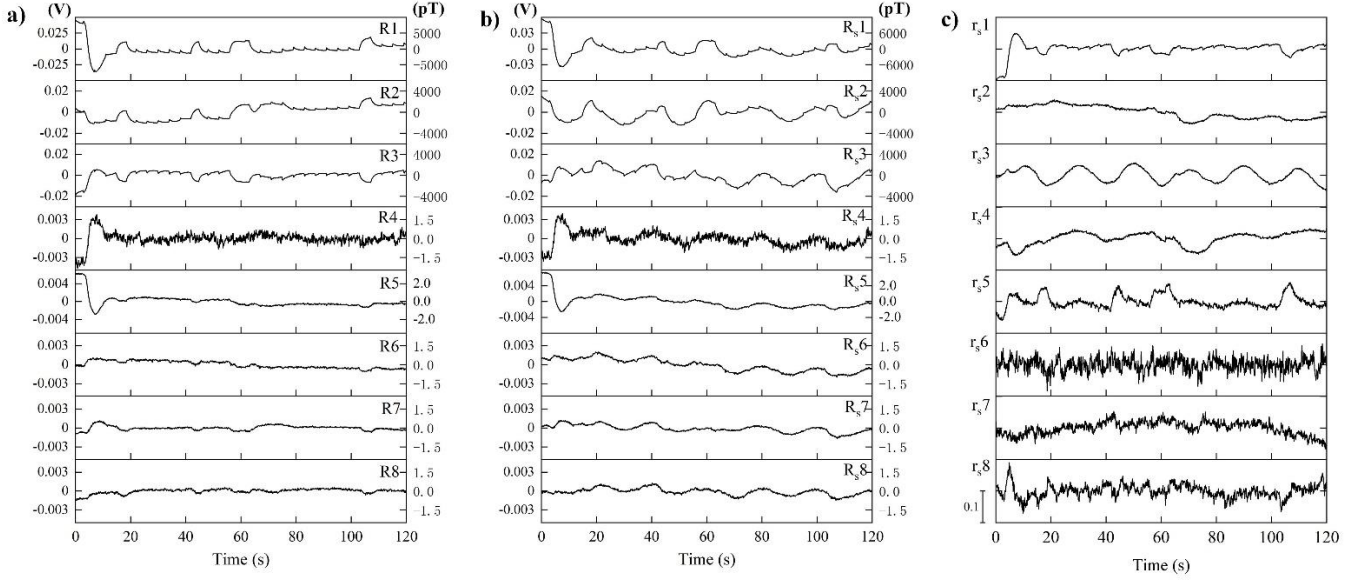

**Fig. 5.** (a) The corresponding output of eight reference channels. (b) The simulated outputs of the eight reference channels $R_s1, \ldots, R_s8$. (c) The signal-noise separation signals using SOBI of the reference channels.

The corresponding output of the three magnetometers and the five one-order tensor gradiometers for the real environment without the measured person are shown in Fig. 5(a). R1, R2, R3 are the magnetometer outputs and the R4, …, R8 are the tensor gradiometer output. Since magnetometers are more sensitive than the first order gradiometers, the output intensity of magnetometers is about an order of magnitude higher than that of the gradiometer in our measurement system.

According to $x_s(t)$, the simulated magnetometer ($R_s1$, $R_s2$, $R_s3$), the simulated tensor gradiometer ($R_s4$, …, $R_s8$) are given by the following expression:

$$(R_s1, R_s2, R_s3) = (R1, R2, R3) + x_s(t) \quad (34)$$

$$(R_s4, \cdots, R_s8) = (R4, \cdots, R8) + 0.1 x_s(t) \quad (35)$$

Fig. 5(b) is simulated reference channel outputs and the signal component influence can be seen.

A set of K = 10 lags was used for SOBI and the maximum time lag is τ = 1 s.

Fig. 5(c) are the eight separated reference signals by using SOBI and the corresponding AR-PSD are shown in Fig. 6. It is clear that $r_s2$ is the separated noise with the strongest AR-PSD peak at 0.3 cpm and $r_s3$ is the separated signal with the strongest peak at 3.0 cpm.

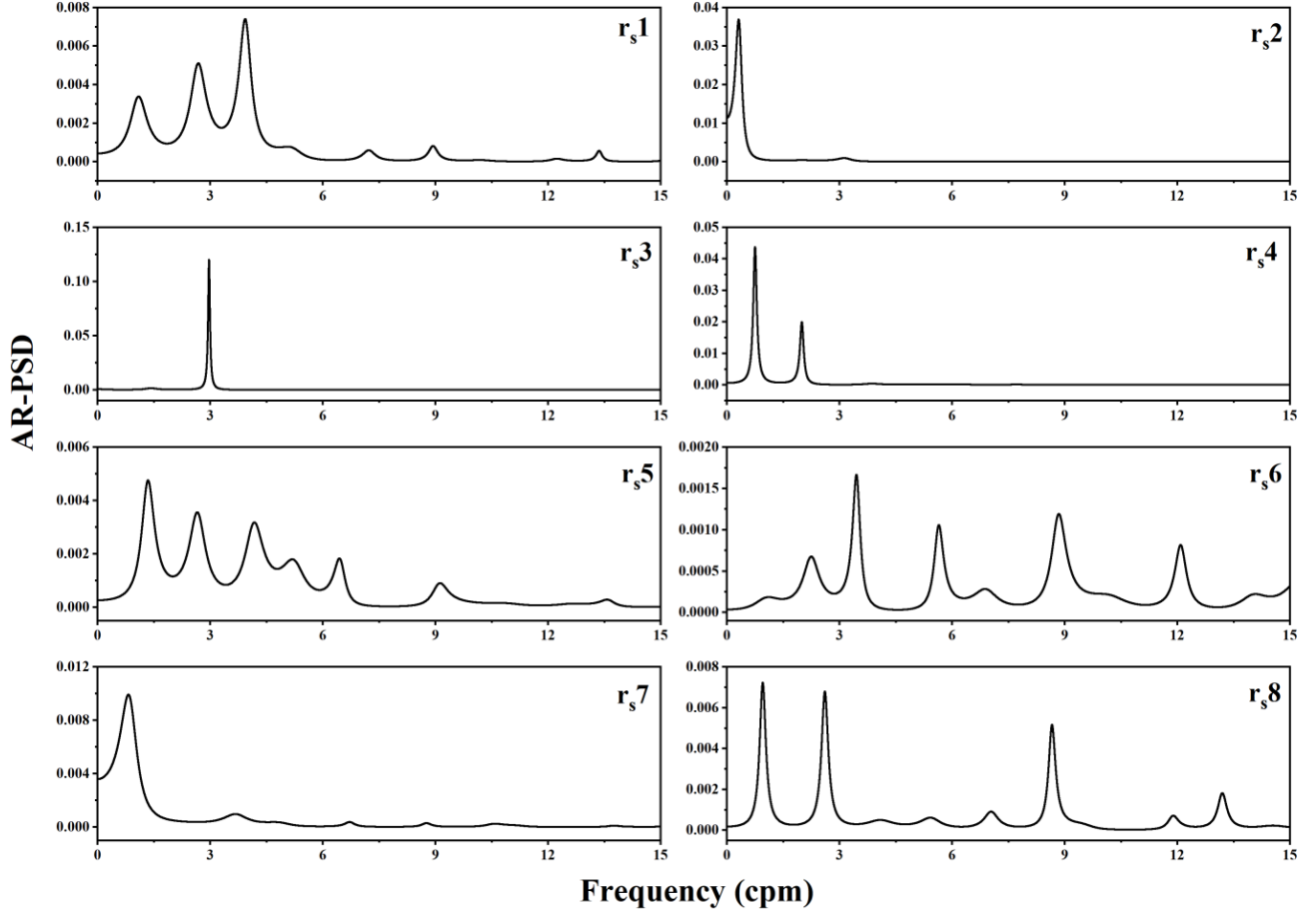

**Fig. 6.** The AR-PSD for the separated reference signals.

TABLE I
THE SEPARATED REFERENCE SNR STATISTICAL TABLE BASED ON THE AR-PSD

| Reference signals after SOBI | Signal (~3 cpm) | Noise (~0.3 cpm) | SNR / dB |
|---|---|---|---|
| $r_s1$ | 0.0051 | 0.00051 | 20 |
| $r_s2$ | 0.00094 | 0.037 | -31.9 |
| $r_s3$ | 0.12 | 0.00034 | 50.9 |
| $r_s4$ | 0.0001 | 0.0009 | -19.1 |
| $r_s5$ | 0.0015 | 0.00029 | 14.3 |
| $r_s6$ | 0.00025 | 0.00004 | 15.9 |
| $r_s7$ | 0.00043 | 0.0042 | -19.8 |
| $r_s8$ | 0.0006 | 0.00023 | 8.3 |

According to the simulation settings, the signal frequency is 3 cpm and the noise frequency is 0.3 cpm. The AR-PSD amplitudes at 3 cpm and 0.3 cpm are counted and the SNR of the reference channel signals after SOBI is calculated (see Table I). The SNR of $r_s2$ is -31.9 dB which is the smallest SNR among the separated reference signals. Therefore, $r_s2$ is selected to be the reference input signal of the SNSWF.

*B. The classic Wiener filter and SNSWF denoising results for simulated MGG*

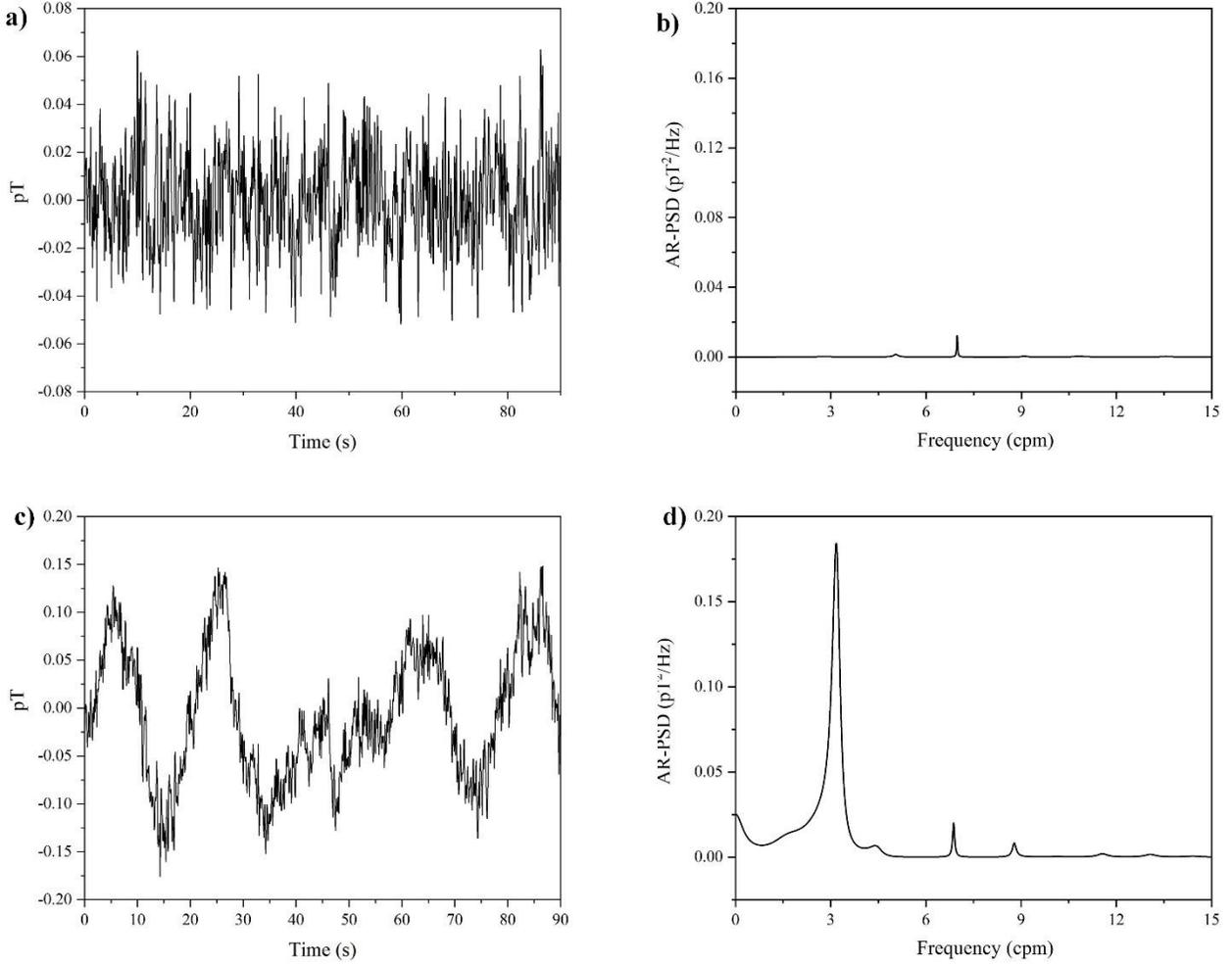

**Fig. 7.** (a) The output of classic Wiener filter. (b) AR-PSD for classic Wiener filter output. (c) The simulated MGG signal after SNSWF. (d) The AR-PSD for SNSWF output.

Fig. 7(a) is the output of classic Wiener filter which directly uses the eight reference channels to filter the signal channel.

Fig. 7(b) is the corresponding AR-PSD for Fig. 7(a) and there is only a small signal peak 2.8 cpm with amplitude 0.00024 pT$^2$/Hz. The introduction of the signal in the reference channels results in poor filtering performance of classical Wiener filters.

Fig. 7(c) is simulated MGG signal after SNSWF and the AR-PSD is shown in Fig. 7(d). The signal peak is about 3.1 cpm and the strength of AR-PSD is 0.18 pT$^2$/Hz. The noise peak is 0.3 cpm with strength 0.014 pT$^2$/Hz. The SNR of the SNSWF output is about 22.2 dB.

V. APPLICATION TO MGG DENOISING

*A. The collected raw data and the signal-noise separation results*

The data sample rate is 1 KHz and then down sampled to 20 Hz.

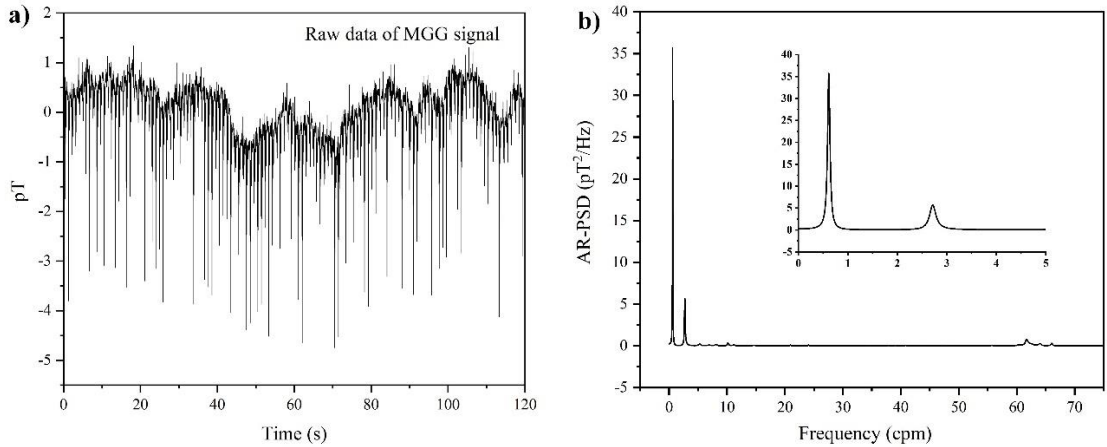

**Fig. 8.** (a) The output of second-order axial gradiometer. (b) AR power spectrum distribution for MGG raw signal.

Fig. 8(a) is the raw data of MGG signal which is measured above the adult stomach within a magnetic shielding room (MSR). Although they are heavily corrupted by noises, the signals amplitude is about 5 pT and the MGG SWF can be found in the time-domine waves which can be recognized in Fig. 8(a).

Fig. 8(b) is the AR power spectrum distribution for MGG raw signal. The MGG SWF is 2.7 cpm and the strength of AR-PSD is about 5.63 $pT^2$/Hz. The strongest noise peak is 0.6 cpm with strength 35.7 $pT^2$/Hz. Therefore, the SNR of the raw MGG signal is about -16.0 dB.

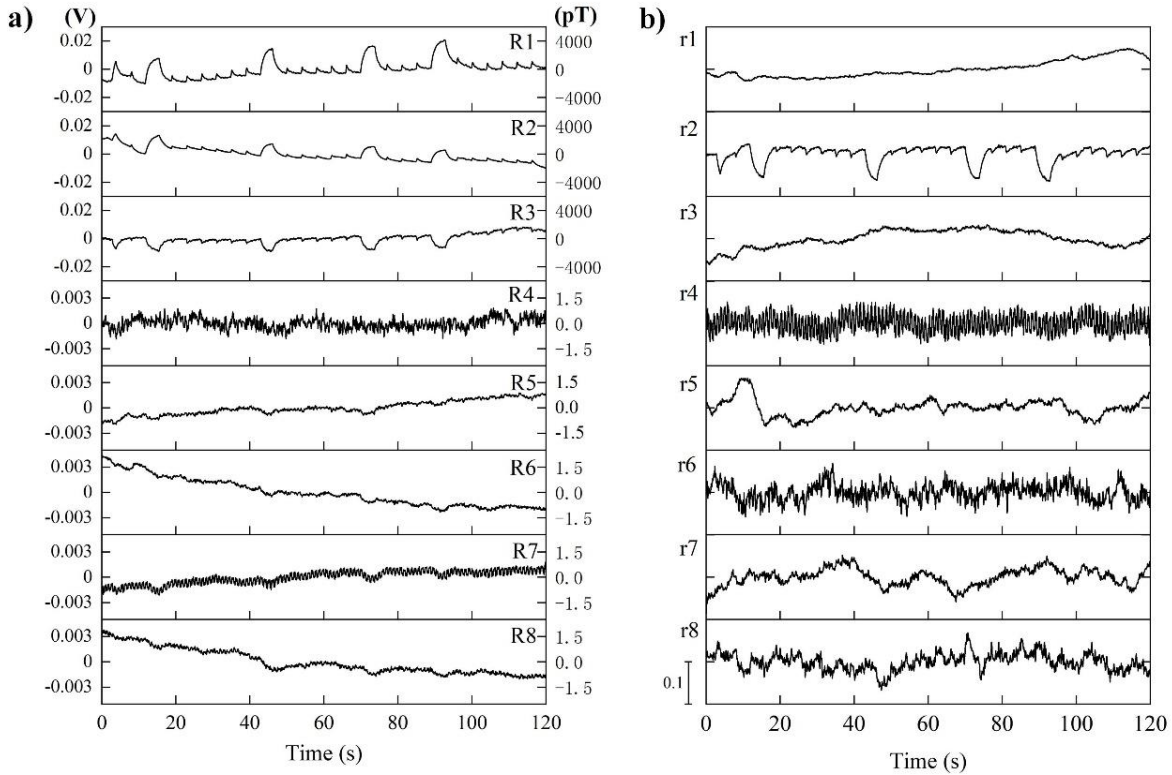

**Fig. 9.** (a) The output of eight reference channel signals. (b) The signal-noise separation signals using SOBI.

Fig. 9(a) are the output voltage signals of reference channels and the length of time for data collection is 120 s. R1, R2 and R3 are the output of three magnetometers and the signal amplitude is about 40 mV and 8 nT. They usually measure the three zero-order components of the environmental field, which show noticeable low-frequency interference noises. R4 to R8 are the output of the five first order gradiometers and the signal amplitude is about 6 mV and 3 pT. They are used to measure the five first-order components of the environmental field. The low-frequency interference noises are clear shown in R4, R5, R6 and R8. These reference channel outputs are typically used directly as reference input signals in the classic Wiener filter, which the filter effects are shown in Fig. 11(a) and Fig. 20 (b).

Fig. 9(b) are the signal-noise separation signals of reference channels after SOBI operation. r1 to r8 separately represent one feature component and the different noise sources are separated from each other.

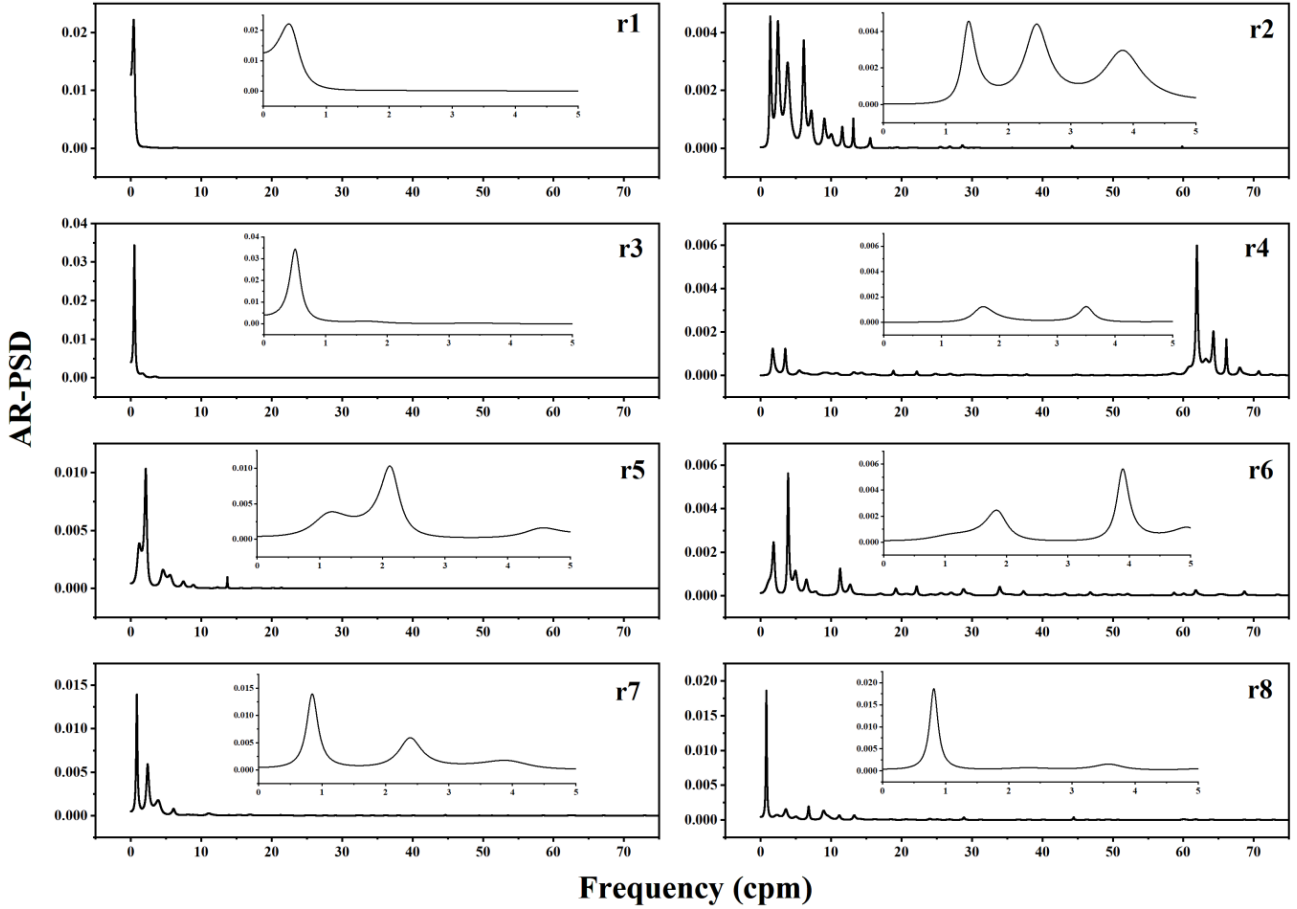

**Fig. 10.** The AR-PSD for the separated reference signals.

TABLE II
SNR AND MAIN SEPARATED NOISE PEAKS STATISTICAL TABLE

| Reference signals after SOBI | SNR / dB | Main Noise Peaks (≤ 2.5 cpm or ≥ 3.5 cpm) / cpm |
|---|---|---|
| r1 | -42.5 | 0.4 |
| r2 | -0.6 | 1.4 |
| r3 | -39.2 | 0.5 |
| r4 | -13.7 | 61.9 |
| r5 | -16.1 | 2.1 |
| r6 | -22.5 | 3.9 |
| r7 | -9.2 | 0.8 |
| r8 | -22.2 | 0.8 |

Table II is the SNR and main separated noise peaks statistical results according to the AR-PSD. In order to compare the reference signals after SOBI, SNR and the main noise peak frequency are considered for choosing the reference input signals. The signal peak which is chosen from frequencies range within $3 \pm 0.5$ cpm and the strongest noise (≤ 2.5 cpm or ≥ 3.5 cpm) peak are used for SNR calculation. The lower SNR, the better it is to be used as the reference input signal. Besides the strong low frequency noise, it is better to consider the relative high frequency interference such as heartbeat.

The AR-PSD of r1 and r3 are both remain the low frequency part shown in Fig. 10, but the SNR and the peak frequency of r1 is lower than r3. In order to denoise the low frequency interference more efficiently, r1 is used as one reference input for SNSWF.

Although the SNR of r4 (-13.7 dB) is not the lowest, it mainly retains the high frequency (61.9 cpm) noise part from heart beat interference noise shown in Fig. 10. Since its frequency is relatively far from MGG signal, it can be also used as the reference input for SNSWF. While the peak frequency of other parts (r2, r5, r6, r7, r8) contain the signal frequencies ($3 \pm 0.5$ cpm), they are not used as the SNSWF input.

*C. The classic Wiener filter and SNSWF denoising results for MGG*

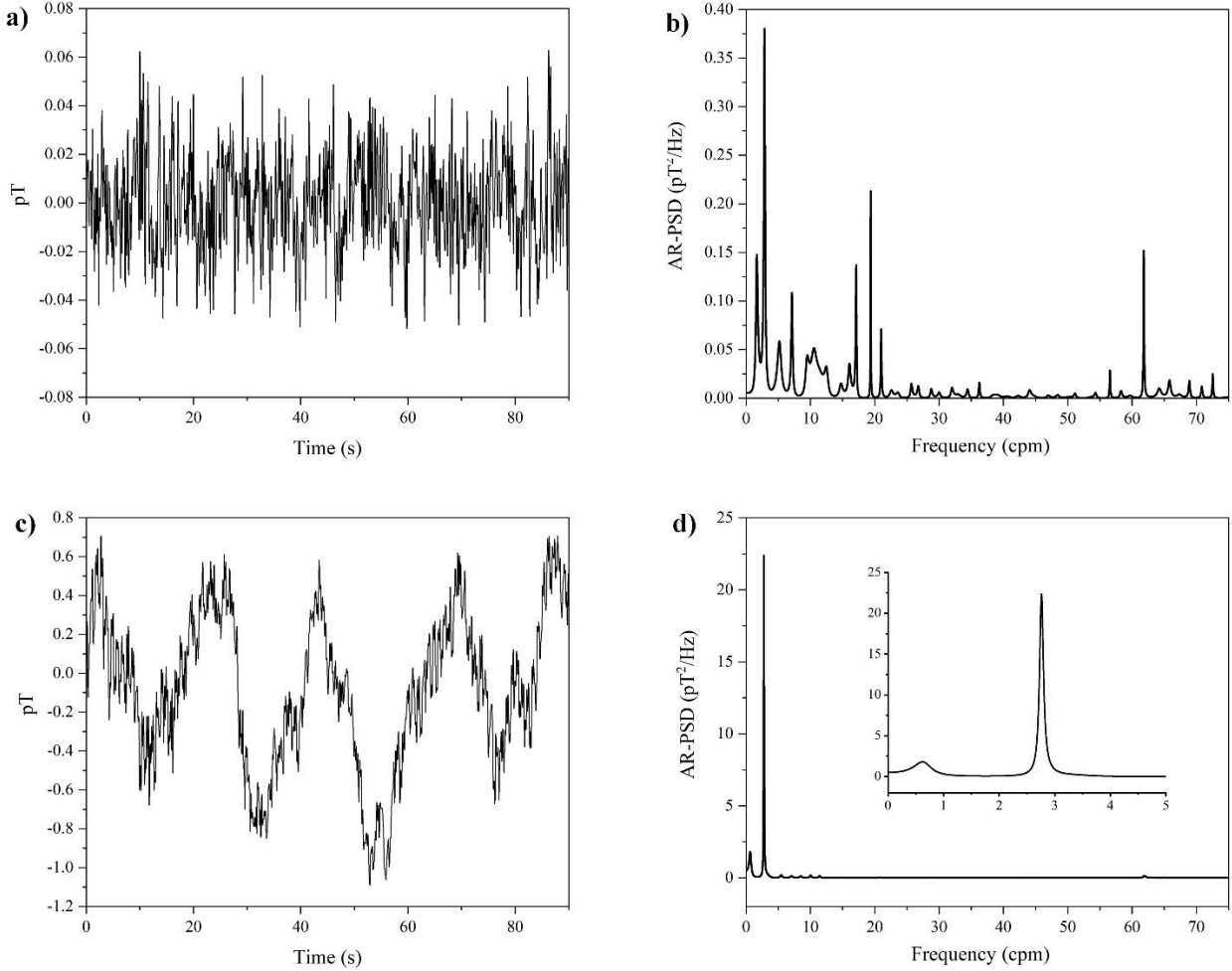

**Fig. 11.** (a) The output of classic Wiener filter. (b) AR power spectrum distribution for classic Wiener filter output. (c) MGG signal after SNSWF. (d) The AR-PSD for SNSWF output.

Fig. 11(a) is the output of Wiener filter shown as Fig. 2 that the outputs of signal channel and reference channels are directly used as the primary input and the reference input. It is not clear to find the time-domain MGG slow waves compared with Fig. 8(a), which can be caused by ill-conditioned solutions of the Wiener filters.

Fig. 11(b) is the AR power spectrum distribution (AR-PSD) of the output. One peak around 3 cpm is 2.8 cpm and the strength of AR-PSD is about 0.38 $pT^2/Hz$. There are many noise peaks among the frequency range from 12 cpm to 20 cpm, which can be other noise sources such as the respiration interference. Considering that the strongest noise peak is 19.3 cpm with strength 0.21 $pT^2/Hz$, the SNR of the output is about 5.2 dB. One another peak is about 61.8 cpm with strength 0.15 $pT^2/Hz$.

After signal-noise separation, r1 and r4 are used as the reference inputs of SNSWF. The SNSWF denoising effects are shown in Fig. 11(c) and Fig. 11(d).

Fig. 11(c) is the output of SNSWF. It is clear to see the SWF from the time-domain MGG waveforms compared with Fig. 11(a).

Fig. 11(d) is the MGG signal AR-PSD after SNSWF. The peak is about 2.8 cpm and the strength of AR-PSD is 22.37 $pT^2/Hz$. The other peak is 0.6 cpm with strength 1.80 $pT^2/Hz$. The SNR of the SNSWF output is about 21.9 dB.

After the SNSWF, the filter SNR is 16.7 dB better than the classic Wiener filter.

## VI. CONCLUSION

In this paper, we proposed a new signal processing framework called signal-noise separation based Wiener filter (SNSWF). It separates the main noise as the input signal of the filter to improve the output SNR of Wiener filter. We also introduced a high-pass Wiener filter with adjustable cutoff frequency. The filter was successfully applied to the noise suppression for MGG signal detection. Using the SNSWF, the filter SNR is 16.7 dB better than the classic Wiener filter. The proposed SNSWF approach can be adopted for many signal processing applications, especially for biomagnetic application fields such as magnetoencephalography (MEG), in the future.

## REFERENCES

[1] S. Somarajan, N. D. Muszynski, D. Hawrami, J. D. Olson, L. K. Cheng and L. A. Bradshaw, "Noninvasive Magnetogastrography Detects Erythromycin-Induced Effects on the Gastric Slow Wave," in IEEE Transactions on Biomedical Engineering, vol. 66, no. 2, pp. 327-334, Feb. 2019, doi: 10.1109/TBME.2018.2837647.

[2] L. A. Bradshaw, J. K. Ladipo, D. J. Staton, J. P. Wikswo and W. O. Richards, "The human vector magnetogastrogram and magnetoenterogram," in IEEE Transactions on Biomedical Engineering, vol. 46, no. 8, pp. 959-970, Aug. 1999, doi: 10.1109/10.775406.

[3] J. C. Erickson, C. Obioha, A. Goodale, L. A. Bradshaw and W. O. Richards, "Detection of Small Bowel Slow-Wave Frequencies From Noninvasive Biomagnetic Measurements," in IEEE Transactions on Biomedical Engineering, vol. 56, no. 9, pp. 2181-2189, Sept. 2009, doi: 10.1109/TBME.2009.2024087.

[4] Bradshaw LA, Sims JA, Richards WO. Noninvasive assessment of the effects of glucagon on the gastric slow wave. Am J Physiol Gastrointest Liver Physiol. 2007 Nov;293(5):G1029-38. doi: 10.1152/ajpgi.00054.2007.

[5] S. Harmeling, A. Ziehe, M. Kawanabe and K. Müller, "Kernel-Based Nonlinear Blind Source Separation," in Neural Computation, vol. 15, no. 5, pp. 1089-1124, 1 May 2003, doi: 10.1162/089976603765202677.

[6] D. Martinez and A. Bray, "Nonlinear blind source separation using kernels," in IEEE Transactions on Neural Networks, vol. 14, no. 1, pp. 228-235, Jan. 2003, doi: 10.1109/TNN.2002.806624.

[7] Uusitalo, M.A., Ilmoniemi, R.J. Signal-space projection method for separating MEG or EEG into components. Med. Biol. Eng. Comput. 35, 135–140 (1997). https://doi.org/10.1007/BF02534144.

[8] A. Hyvärinen and E. Oja, "A Fast Fixed-Point Algorithm for Independent Component Analysis," in Neural Computation, vol. 9, no. 7, pp. 1483-1492, 10 July 1997, doi: 10.1162/neco.1997.9.7.1483.

[9] B. Schölkopf, A. Smola and K. Müller, "Nonlinear Component Analysis as a Kernel Eigenvalue Problem," in Neural Computation, vol. 10, no. 5, pp. 1299-1319, 1 July 1998, doi: 10.1162/089976698300017467.

[10] S. Y. Chang and H. -C. Wu, "Tensor Wiener Filter," in IEEE Transactions on Signal Processing, vol. 70, pp. 410-422, 2022, doi: 10.1109/TSP.2022.3140722.

[11] B. Widrow et al., "Adaptive noise cancelling: Principles and applications," in Proceedings of the IEEE, vol. 63, no. 12, pp. 1692-1716, Dec. 1975, doi: 10.1109/PROC.1975.10036.

[12] I. Constantin, C. Richard, R. Lengelle and L. Soufflet, "Nonlinear Regularized Wiener Filtering With Kernels: Application in Denoising MEG Data Corrupted by ECG," in IEEE Transactions on Signal Processing, vol. 54, no. 12, pp. 4796-4806, Dec. 2006, doi: 10.1109/TSP.2006.882115.

[13] A. Belouchrani, K. Abed-Meraim, J. -F. Cardoso and E. Moulines, "A blind source separation technique using second-order statistics," in IEEE Transactions on Signal Processing, vol. 45, no. 2, pp. 434-444, Feb. 1997, doi: 10.1109/78.554307.

[14] H. Li, S. Zhang, C. Zhang and X. Xie, "SQUID-Based MCG Measurement Using a Full-Tensor Compensation Technique in an Urban Hospital Environment," in IEEE Transactions on Applied Superconductivity, vol. 26, no. 6, pp. 1-5, Sept. 2016, Art no. 1601805, doi: 10.1109/TASC.2016.2569507.